\documentclass[aps, nofootinbib,twocolumn]{revtex4}

\usepackage{amsmath}
\allowdisplaybreaks

\usepackage{makeidx}
\usepackage{amsfonts}
\usepackage[usenames,dvipsnames]{pstricks}
\usepackage{subfigure}
\usepackage{epsfig}
\usepackage{pst-grad} 
\usepackage{mathtools}
\usepackage[colorlinks,hyperindex]{hyperref}
\usepackage{array}

\hypersetup
{	colorlinks,%
	citecolor=black,%
	linkcolor=black,%
	urlcolor=black,%
}

\usepackage{allpurpose}

\usepackage{pdfpages} 

\newcommand\nn{{\nonumber}}

\begin{document}

\title{Perturbative deflection angle for signal with finite distance and general velocities}

\author{Ke Huang}
\affiliation{Center for Astrophysics \& MOE Key Laboratory of Artificial Micro- and Nano-structures, School of Physics and Technology, Wuhan University, Wuhan, 430072, China}

\author{Junji Jia}
\email{Corresponding author: junjijia@whu.edu.cn}
\affiliation{Center for Astrophysics \& MOE Key Laboratory of Artificial Micro- and Nano-structures, School of Physics and Technology, Wuhan University, Wuhan, 430072, China}

\date{\today}

\begin{abstract}
We propose a perturbative method to compute the deflection angle of both null and massive particles for source and detector at finite distance. This method applies universally to the motion of particles with general velocity in the equatorial plane of stationary axisymmetric spacetimes or static spherical symmetric spacetimes that are asymptotically flat. The resultant deflection angle automatically arranges into a quasi-inverse series form of the impact parameter, with coefficients depending on the metric functions, the signal velocity and the source and detector locations through the apparent angles. In the large impact parameter limit, the series coefficients are reduced to rational functions of sine/cosine functions of the zero order apparent angle.
\end{abstract}


\maketitle

\section{Introduction}
One hundred years ago, Eddington's observation of the lightray bending helped establishing General Relativity (GR) as the correct description of gravity \cite{Dyson:1920cwa}. Nowadays, the deflection of lightrays has become the foundation of gravitational lensing (GL), an important observational tool for astrophysics and cosmology. It is used to study
properties of the supernova \cite{Sharon:2014ija},
coevolution of supermassive black holes (BHs) and galaxies \cite{Peng:2006ew},
mass distributions of galaxy clusters \cite{Bartelmann:1999yn} and the Universe  \cite{Ade:2015zua},
cosmological parameters \cite{Refregier:2003ct,Lewis:2006fu} and dark matter and energy \cite{Metcalf:2001ap, Hoekstra:2008db}.

Traditionally, lightrays have been the main messengers in the observation of trajectory bending and GL. However, with the observation of extragalactic neutrinos from SN1987A \cite{Hirata:1987hu, Bionta:1987qt} and blazar TXS 0506+056 \cite{IceCube:2018dnn,IceCube:2018cha}, and the more recent  gravitational waves (GWs) \cite{Abbott:2016blz,Abbott:2016nmj,Abbott:2017oio,TheLIGOScientific:2017qsa,Monitor:2017mdv},  it is now clear that these two kinds of signals can also act as astrophysical messengers and experience deflection and potentially be observed in lensing scenarios \cite{barrow1987lensing,  Mena:2006ym, Eiroa:2008ks,  Fan:2016swi,Wei:2017emo,Yang:2018bdf}. Neutrinos are known to have nonzero masses  \cite{Tanabashi:2018oca} and GWs might also have a non-luminal speed in some gravitational theories beyond GR \cite{Sakstein:2017xjx,Baker:2017hug}. In considering GLs of massive particles, the fundamental quantity -- the deflection angle, should be computed along timelike geodesics for source and observers at finite distance. Previously, most works still use the deflection angles computed along null rays for sources and observers at infinity \cite{barrow1987lensing, Mena:2006ym, Eiroa:2008ks, Fan:2016swi,Wei:2017emo,Yang:2018bdf}.

Recently, deflection angles of timelike particles with arbitrary velocity were considered for some particular spacetimes using the elliptical integrals of the geodesic equations and their correlation with the neutrino absolute mass and mass order were proposed \cite{Jia:2015zon,Pang:2018jpm}. Meanwhile, the deflection angle for sources and detectors at finite distance has also been considered using the Gauss-Bonnet theorem method \cite{Gibbons:2008rj,Ishihara:2016vdc,Arakida:2017hrm,Crisnejo:2018uyn,Jusufi:2018jof, Ovgun:2018tua,Kumaran:2019qqp,Zhu:2019ura,Ovgun:2019qzc,Javed:2019kon,Li:2019qyb}.
(see \cite{Ono:2019hkw} and references therein for a partial review). However, a general and yet simple method that works for arbitrary static spherically symmetric (SSS) spacetimes, and equatorial motion in stationary axisymmetric (SAS) spacetimes, is still lacking. In this work, we report a perturbative method to compute the deflection angle in these cases for signal with arbitrary velocity and from sources and to detectors at arbitrary distances. We use the geometric unit $G=c=1$ throughout the paper.

\section{Deflection angle in SAS spacetimes}
The most general SAS metric can be described by \cite{Sloane:1978ne,Ono:2017pie}
\be
   \dd s^2=-A\dd t^2+B \dd t\dd \phi+C\dd\phi^2+D\dd x_1^2 +F\dd x_2^2 \label{eq:sasgm}
\ee
where $(t,~\phi,~x_1,~x_2)$ are the coordinates and $A,~B,~C,~D$ and $F$ are functions of $x_1$ and $x_2$ only. The $x_1,~x_2$ are often chosen as cylindrical coordinates $(\rho,~z)$ such as in the Weyl-Lewis-Papapetrou line element \cite{Sloane:1978ne}, or equivalently the spherical coordinates $(r,~\theta)$ such as the Boyer-Lindquist coordinates of the Kerr-Newman (KN) spacetime. Here we choose the latter because they
allow a straightforward reduction to the deflection angle in the SSS spacetime by simply setting $B(r)=0$ in Eq. \eqref{eq:sasgm}.
Furthermore, we assume that the spacetime permits motion of particles in a plane with fixed $\theta$, which can always be shifted to $\theta=\pi/2$ and called the equatorial plane henceforth. Motion in this plane then effectively takes place in a 1+2 dimensional spacetime, whose metric after suppressing the $\theta$ coordinate becomes
\be
\dd s^2=-A(r)\dd t^2+B(r)\dd t\dd \phi +C(r)\dd \phi^2+D(r)\dd r^2. \label{metric2}
\ee

It is then routine to find the geodesic equations for equatorial motion in this metric and the corresponding first integrals and the equation of motion for $ \dd\phi/\dd r$. Consequently, the deflection angle for a ray from source at $r_s$ to a detector at $r_d$ is \cite{Jia:2020dap}
\bea
\alpha&=&
\lsb \int_{r_0}^{r_s}+\int_{r_0}^{r_d}\rsb\dd r \frac{\sqrt{AD}}{\sqrt{AC+B^2/4}}\nn\\
&&\times\frac{(2LA -EB )}{\sqrt{\lb 4AC+B^2\rb \lb E^2-\kappa A\rb-( 2LA -EB)^2}}\nn\\
&& +\beta_s+\beta_d-(2i+1)\pi \label{asaiint}\\
&\equiv &\Delta \phi +\beta_s+\beta_d-(2i+1)\pi. \label{eq:dphidef}
\eea
Here the first term in Eq. \eqref{asaiint} is the change of the angular coordinate from angular coordinate $\phi_s$ of the source to $\phi_d$ of the detector, i.e., $\Delta \phi=\phi_d-\phi_s$.
$r_0$ in $\Delta\phi$ is the minimal $r$ at which the rays turn from inward to outward motion in the radial direction. $L$ and $E$ are respectively the angular momentum and energy of the unit mass of the particle. $\beta_{s,d}$ are the small apparent angles of the signal measured locally against the radial direction at the position of the source and detector, respectively. They are present in Eq. \eqref{eq:dphidef} to guarantee that the deflection angle is geometrically invariant \cite{Ishihara:2016vdc,Ono:2019hkw} and recovers zero in the finite distance case in the Minkowski spacetime limit. The $-(2i+1)\pi$ term is to modulate away the necessary part such that $\alpha$ is between $-\pi$ and $\pi$.

In asymptotically flat spacetimes, on which this work concentrates, $L$ and $E$ are related to the velocity $v$ at infinity and impact parameter $b$ by
\be
|L|=|\mathbf{p}\times\mathbf{r}| =\frac{v}{\sqrt{1-v^2}}b,~ E=\frac{1}{\sqrt{1-v^2}}. \label{eq:ledef}\ee
The angular momentum $L$ can also be related to $r_0$ using the radial geodesic equation $\dd r/\dd t|_{r=r_0}=0$, to find
\be
L=\frac{-s \sqrt{\lsb 4A(r_0)C(r_0)+B(r_0)^2\rsb\lsb E^2-\kappa A(r_0)\rsb}+EB(r_0)}{2A(r_0)}, \label{lsasdef}
\ee
where sign $s=\pm 1$. In the small spacetime spin limit, $|B(r)|\ll 1$, the first term in the numerator dominates the second, and therefore $s=+1$ and $-1$ correspond to the cases that the particle motion is retrograde and prograde respectively.

Further using Eqs. \eqref{eq:ledef} and \eqref{lsasdef}, one can establish the following relation between the impact parameter $b$ and $r_0$
\bea
\frac{1}{b}&=&\frac{2 A(r_0)\sqrt{E^2-\kappa}+sB(r_0)E/b}{s\sqrt{\lsb 4A(r_0)C(r_0)+B(r_0)^2\rsb\lsb E^2-\kappa A(r_0)\rsb}} \nn\\
&\equiv & p\lb b, \frac{1}{r_0}\rb. \label{eq:pdef}
\eea
In the last step we denoted $1/b$ on the left-hand side as a function $p$ of both $b$ and $1/r_0$. With this function, it is straightforward to prove using the projection of the signal's four velocity
\bea
\frac{\partial x^{\mu}}{\partial \tau} &=&\left[\frac{EC+LB/2}{AC+B^2/4},\sqrt{\frac{E^2C+ELB-L^2A}{(AC+B^2/4)D}-\frac\kappa D},\right.\nn\\
&&\left.0,\frac{LA-EB/2}{AC+B^2/4}\right]
\eea
onto the four velocities of the source and detector $\partial x^{\mu}/\partial \tau^\prime=(1/\sqrt{A(r_{s,d})},0,0,0) $ \cite{bk:hobson} that the apparent angles $\beta_{s,d}$
are simply
\be
\beta_{s,d}=\arcsin\lsb b\cdot p(b,1/r_{s,d})\rsb.
\label{eq:betadef}\ee

We now propose a special change of variable in the deflection angle formula \eqref{asaiint}, following which we can do a series expansion of the new variable and then prove rigorously its integrability to any desired order of $1/b$ for signals with arbitrary velocity in general SAS metrics and from sources and to detectors at arbitrary distance.
The change of variable is inspired by the function $p(b,x)$ defined in Eq. \eqref{eq:pdef}. We can first formally obtain $p(b,x)$'s inverse function $q(b,x)$ with respect to its second argument such that
\be \frac{1}{r_0}=q\lb b,~\frac{1}{b}\rb. \label{eq:qdef}\ee
Then making a change of variable in Eq. \eqref{asaiint} from $r$ to $u$, who are connected by the relation
\be
\frac{1}{r}=q\lb b, \frac{u}{b}\rb, \label{eq:udef}\ee
we see that after using Eqs. \eqref{eq:pdef}, \eqref{eq:qdef} and \eqref{eq:udef} repeatedly and after some tedious but still element algebra, the integral limits, the first and second factors and the measure in Eq. \eqref{asaiint} become respectively
\begin{align}
&r_0\to 1,~r_{s,d}\to b\cdot p\lb b, \frac{1}{r_{s,d}}\rb=\sin\beta_{s,d},\nn \\
&\sqrt{\frac{A(r)D(r)}{A(r)C(r)+B(r)^2/4}} \to \sqrt{\frac{A(1/q)D(1/q)}{A(1/q)C(1/q)+B(1/q)^2/4}}  ,\nn \\
&\frac{\lsb 2LA(r)-EB (r)\rsb } {\sqrt{\lsb 4A(r)C(r)+B(r)^2\rsb \lsb E^2-\kappa A (r)\rsb-\lsb 2LA(r)- EB (r) \rsb^2}}\nn\\
&\hspace{4cm}\to \frac{u}{\sqrt{1-u^2}}, \nn\\
&\dd r \to -\frac{1}{p_2\lb b, q\rb q^2}\frac{1}{b}\dd u, \nn
\end{align}
where $p_2$ is the derivative of function $p(b,q)$ in Eq. \eqref{eq:pdef} with respect to its second argument $q=q(b,u/b)$.
Collecting these together, the change of the angular coordinate in Eq. \eqref{asaiint} becomes
\bea
\Delta\phi&=&\lsb \int_{\sin\beta_s}^1+\int_{\sin\beta_d }^1\rsb y\lb b, \frac{u}{b}\rb  \frac{\dd u}{\sqrt{1-u^2}} \label{eq:idef3}
\eea
where
\be y\lb b, \frac{u}{b}\rb =
\sqrt{\frac{A(1/q)D(1/q)}{A(1/q)C(1/q)+B(1/q)^2/4}}\frac{1}{p_2\lb b, q\rb q^2}
\frac{u}{b} .\label{eq:ydef}\ee

The key is to note that this function $\displaystyle  y\lb b, \frac{u}{b}\rb$ depends on $u$ only through the ratio $\frac{u}{b}$, either directly or through the function $q(b,u/b)$. It then can be expended with respect to its second argument
\be
y\lb b, \frac{u}{b}\rb=\sum_{n=0}^\infty y_n(b)\lb \frac{u}{b}\rb^n, \label{eq:yexp}
\ee
where $y_n$ are the coefficients expressible as polynomials of asymptotic expansion coefficients of the metric functions.
Substituting this into the deflection angle \eqref{eq:idef3} and do a simple change of variable $u=\sin\xi$ suggested by the denominator $\sqrt{1-u^2}$,
this becomes
\be
\Delta\phi=\sum_{n=0}^\infty \frac{y_n(b)}{b^n} \lsb \int_{\beta_s}^{\frac{\pi}{2}} +\int_{\beta_d}^{\frac{\pi}{2}} \rsb \sin^n \xi\dd\xi
. \label{eq:idef4}
\ee
At this point, the integrability of Eq. \eqref{eq:idef4} to any desired order of $y_n(b)/b^n$ becomes clear because the integral can always be carried out to yield \cite{bk:inttable}
\bea
&&l_n(\beta_s,~\beta_d)\equiv \lsb \int_{\beta_s}^{\frac{\pi}{2}} +\int_{\beta_d}^{\frac{\pi}{2}} \rsb\sin^n \xi\dd \xi=\sum_{i=s,d} \frac{(n-1)!!}{n!!}\times \nn\\
&&\begin{cases}
\displaystyle  \left(\frac{\pi}{2}-\beta_i
    +\cos\beta_i\sum_{j=1}^{[n/2]} \frac{(2j-2)!!} {(2j-1)!!}\sin^{2j-1} \beta_i\right),~n=2k,\\
\displaystyle \cos\beta_i \left(1
    +\sum_{j=1}^{[n/2]} \frac{(2j-1)!!}{(2j)!!} \sin^{2j}\beta_i\right),~n=2k+1.
\end{cases}
\label{eq:cndef}
\eea
This results in a change of the angular coordinate
\be
\Delta\phi=\sum_{n=0}^\infty l_n(\beta_s,~\beta_d)\frac{y_n(b)}{b^n}. \ee
Note when $r_{s,d}$ are set to infinity in asymptotically flat spacetimes, $\beta_{s,d}$ in all $l_n$ would vanish and the $l_n$s become independent of $r_{s,d}$. Besides, from Eq. \eqref{eq:cndef} we see that $l_0=\pi-\beta_s-\beta_d$, which cancels exactly the apparent angles $\beta_{s,d}$ and one $\pi$ in Eq. \eqref{eq:dphidef}. The final result of the deflection angle becomes
\be
\alpha=\sum_{n=1}^\infty l_n(\beta_s,~\beta_d)\frac{y_n(b)}{b^n}
-2i \pi. \label{eq:alphares}
\ee

A few remarks are in order. First, in the above procedure from definition \eqref{asaiint} to result \eqref{eq:alphares}, once the metric functions are given, the only point that one might suspect a difficulty is to find the inverse function $q(b,x)$, because the inverse function of a given function might not always be solvable. Let us point out that here, this is not an issue because what is needed in Eq. \eqref{eq:yexp} or \eqref{eq:idef4} is only the {\it expansion} of $q(b,x)$ but not itself. Using the Lagrange inversion theorem, we can directly find its expansion without solving $q(b, x)$ explicitly.
Second, the result \eqref{eq:alphares} was obtained, after the expansion of the integrand, in a series form. Therefore, except some critical impact parameter(s) $b_c$ below which the integral \eqref{asaiint} diverges, the exact deflection angle will be finite and the series \eqref{eq:alphares} will converge to that exact value as the sum goes to high enough order.
Third, it is always the case that the spacetimes possess some (at least one) characteristic mass scale, i.e., the ADM mass $m$. And previously, deflection angles are often computed in the weak field limit $b\gg m$. However, because now the result \eqref{eq:alphares} is convergent for all $b$ until $b_c$, the weak field limit is not necessary for its validity, neither is the condition $r_s,~r_d\gg b$.

It is often desirable to express the entire deflection angle \eqref{eq:alphares} as a complete series of $1/b$. Noticing the appearance of $\beta_{s,d},~\cos\beta_{s,d}$ and $\sin\beta_{s,d}$ terms in $l_n(\beta_s,~\beta_d)$ in Eq. \eqref{eq:cndef} and the form of $\beta_{s,d}(b,~r_{s,d})$ in Eq. \eqref{eq:betadef}, it is not hard to convince oneself that in the finite distance case and without making a new expansion for large $b$, $l_n(\beta_s,~\beta_d)$ could never be completely rewritten into a series of $1/b$. On the other hand, we are reluctant to do this large $b$ expansion because if this series is truncated, the result will deviate from the true value of the deflection angle before we reach the $b\to b_c$ limit. With that being said however, one can prove using the method of induction that for asymptotically flat SAS spacetimes, whose metric must satisfy as $r\to\infty$ \cite{Bardeen:1973}
\be
A(r),~rB(r),~C(r)/r^2,~D(r)\to \gamma_{A,B,C,D} +\mathcal{O}(r^{-1}), \label{eq:asycond}
\ee
where $\gamma_{A,B,C,D}$ are some constants, each factor $y_n(b)$ in Eq. \eqref{eq:alphares} will always reduce to polynomials of $1/b$ of degree $n$ or smaller. Moreover, since the deflection angle $\alpha$ and $l_n$ are dimensionless while $b$ is of the dimension of mass, $y_n(b)$ must be of dimension $m^n$. Therefore, we can always have
\be
y_n(b)=m^n\sum_{j=0}^{n} y_{n,j}\lb \frac{m}{b}\rb^j.
\ee
where $y_{n,j}$ are the coefficients of the polynomial.

Thus, one can collect the terms in Eq. \eqref{eq:alphares} merely according to the powers of $m/b$ in the factors $y_n(b)/b^n$, and the result should always be expressible as a quasi-series of $m/b$ with coefficients being linear combinations of $l_n$ and $y_{n,j}$. That is,
\bea
\alpha=\sum_{n=1}^\infty \displaystyle \lb\sum_{j=0}^nl_j y_{j,n-j}\rb\lb \frac{m}{b}\rb^n
-2i\pi .\label{eq:alpharesf}
\eea

\section{Deflection angle in the KN spacetime}
Here we choose to carry out the above procedure for the KN spacetime for its relevance to practical GLs. Application of this procedure to much more complicated metrics are also simple \cite{Duan:2020tsq}, if not trivial. The KN metric on the equatorial plane is given by \cite{Misner:1974qy}
\begin{align}
A&=\Delta_-/r^2,        ~& B&=-2a(2mr-q^2)/r^2,~\nonumber\\
C&=r^2+a^2\Delta_+/r^2, ~& D&= r^2/(a^2+\Delta_-),
\end{align}
where $m,~q,~a=J/m$ are respectively the total mass, total charge and angular momentum per unit mass of the spacetime and $\Delta_\pm(r)=r^2\pm(2 mr-q^2)$.
Substituting them into Eq. \eqref{eq:pdef} we obtain the function $p_K$ for the KN spacetime
\be
    p_K\lb b,\frac{1}{r}\rb
    =\frac{v \Delta_-(r)+a \left(2mr-q^2\right)/b} {(a^2+\Delta_-(r)) \sqrt{\Delta_-(r)(v^2-1)+r^2}}.
\label{eq:pdefkn}\ee
However, this function does not allow a simple and explicit inverse function.
Nevertheless, using this and the metric functions in Eq. \eqref{eq:ydef}, we can formally obtain $y(b,u/b)$. Its expansion \eqref{eq:yexp} then can be worked out explicitly.
Clearly, since the KN metric satisfies the asymptotic conditions \eqref{eq:asycond}, the coefficient $y_n(b)$ of the above expansion should be polynomials of $1/b$.
Further substituting into Eq. \eqref{eq:alphares} and collecting terms according to the power of $m/b$, one obtains the deflection angle in the form of Eq. \eqref{eq:alpharesf}
\begin{equation}
\alpha =\sum_{n=1}^{\infty}z_n \lb\frac mb\rb^n -2i\pi, \label{eq:alphasasinz}
\end{equation}
with the coefficients
\begin{align}
    z_1=
        &
        l_1\left[1+\frac1{v^2}\right] , \label{eq:zsmall}\\
    z_2=
        & \frac{2l_1s\hat{a}}v +
        \frac{ l_2}{2} \left[3+\frac{12}{v^2}-\hat{q}^2 \left(1+
        \frac2{v^2}\right)\right], \nn
\end{align}
and higher orders to $z_9$ have been obtained but not shown for their excessive length (see Supplement Material \cite{suppmat} for these high orders).
Here $\hat{q}\equiv q/m,~\hat{a}\equiv a/m$ and $l_n$ are given in Eq. \eqref{eq:cndef} in which $\beta_{s,d}=\arcsin\lb b\cdot p_K(b,1/r_{s,d})\rb$ according to Eqs. \eqref{eq:betadef} and \eqref{eq:pdefkn}.

\begin{figure}[htp!]
\includegraphics[width=0.45\textwidth]{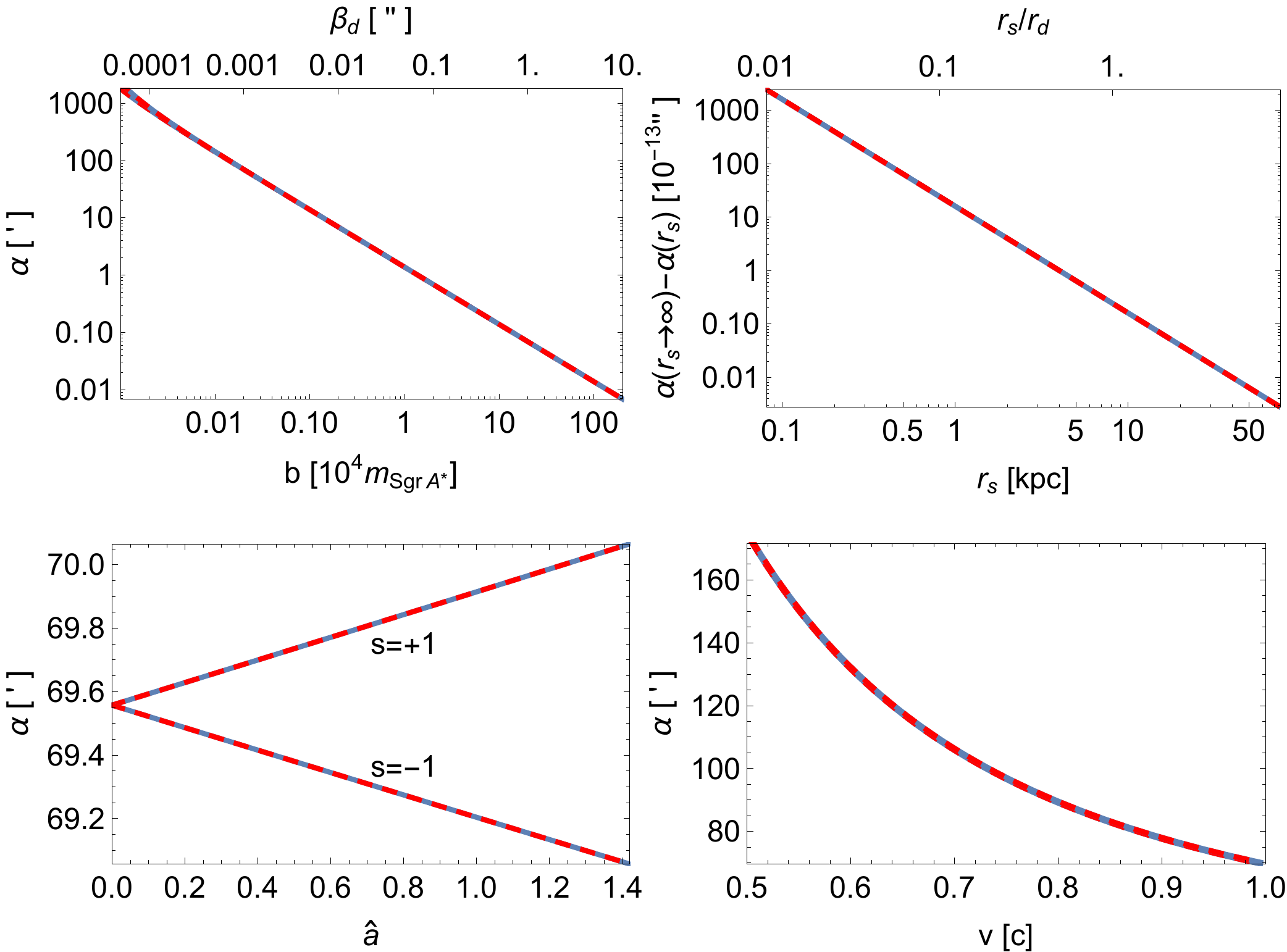}
\caption{\label{fig:knplot} The dependance of the deflection angle $\alpha_K$ of the Sgr A$^*$ BH, on $b$ (upper left), $r_s$ (upper right), $\hat{a}$ (lower left) and $v$ (lower right). The parameters chosen are $m_{\mathrm{Sgr~A}*}=4.12\times 10^6M_\odot$, $q=3\times 10^8$ [C], $\hat{a}=0.71$, $b=200.6m_{\mathrm{Sgr~A}*}$, $r_d=8.12$ [kpc], $r_s=r_d$ and $v=1$ \cite{sgrapara}, except the one that is varied in the $x$-axis of each subplot. Both the prograde ($s=-1$) and retrograde ($s=+1$) cases are plotted but they visually overlap except in the lower left subplot.}
\end{figure}

In Fig. \ref{fig:knplot}, we model the supermassive BH Sgr A* as a KN BH and plot the deflection angle from source in the Galaxy to detector on Earth, using Eq. \eqref{eq:alphasasinz} truncated to $n=9$ as an approximation (grey solid line beneath) and numerical integration of Eq. \eqref{eq:dphidef} as an exact value (dashed red line). It is seen that the analytical result agrees perfectly with the exact value in all plots. The effect of $b$ is as expected while that of finite $r_{s,d}$ is very small. The retrograde motion $(s=+1)$ has a slightly larger deflection angle than the prograde motion $(s=+1)$ for all nonzero $\hat{a}$, including the non-BH case with $\hat{a}>1$. The decrease of the particle velocity increases the deflection angle, also as expected.

\section{Reduction in the SSS spacetimes}
The above method that is valid in the equatorial plane of SAS spacetimes apparently should work for arbitrary geodesic motions in any SSS spacetime because the latter can always be transformed to a motion in the equatorial plane of an SAS spacetime.
Here we present the SSS case because its final result has one important simplification comparing to the SAS case.

The most general SSS spacetime can be described by the metric
\be \dd s^2=-A(r)\dd t^2+D(r)\dd r^2+C(r)(\dd\theta^2+\sin^2\theta\dd \phi^2). \label{sss1} \ee
Although one can always scale $r$ so that $C(r)=r^2$, we will keep $C(r)$ general
until Eq. \eqref{eq:alphalbsss} because for some special coordinate system keeping $C(r)$ general might be more useful. Setting $B=0$ in Eq. \eqref{eq:pdef}, the function $p$ in this case becomes
\be
\frac{1}{b}=\frac{\sqrt{E^2-\kappa}}{\sqrt{E^2-\kappa A(r_0)}}\sqrt{\frac{A(r_0)	}{C(r_0)}}\equiv p\lb \frac{1}{r_0}\rb. \label{eq:pdefs}
\ee
An important difference between this and the corresponding function \eqref{eq:pdef} in the SAS case is that the $p$ here does not depend on $b$ but only on $1/r_0$.

The change of the angular coordinate \eqref{eq:idef3} becomes
\be
\Delta\phi=\lsb \int_{\sin\beta_s}^1+\int_{\sin\beta_d }^1 \rsb y\lb \frac{u}{b}\rb  \frac{\dd u}{\sqrt{1-u^2}} \label{eq:idef3s}
\ee
where \be
\beta_{s,d}=\arcsin\lsb b\cdot p(1/r_{s,d})\rsb \label{eq:betadefs}\ee
with $p$ given in Eq. \eqref{eq:pdefs} and
\be  y\lb \frac{u}{b}\rb =
\sqrt{\frac{D(1/q)}{C(1/q)}} \frac{1}{p^\prime\lb q\rb q^2}\frac{u}{b} ,\ee
in which $q=q(u/b)$ and $q(x)$ is the inverse of $p(x)$  in Eq. \eqref{eq:pdefs}.
The expansion of $y(u/b)$ now becomes
\be
y\lb \frac{u}{b}\rb=\sum_{n=0}^\infty y_n\lb\frac ub\rb^n. \label{eq:sssynexp}
\ee
This is in contrast to Eq. \eqref{eq:yexp} in that $y_n$ here is independent of $b$ due to the fact that $q(u/b)$ and consequently $y(u/b)$ depend on both $u$ and $b$ only through their ratio. Indeed one can prove that to fix the coefficients up to the $n$-th order in Eq. \eqref{eq:sssynexp}, i.e., $y_{1,\cdots,n}$, one only needs to know the coefficients of the asymptotic expansion of the metric functions
\be
    A(r)=1+\sum_{n=1}^{\infty}\frac{a_n}{r^n},
    ~D(r)=1+\sum_{n=1}^{\infty}\frac{d_n}{r^n},
    \frac{C(r)}{r^2}=1+\sum_{n=1}^{\infty}\frac{c_n}{r^n},\label{eq:sssasyexp}
\ee
to the $n$-th order. To the second order, they are
\begin{align}
    y_1=&
        -\frac{a_1}{2 v^2}+\frac{d_1}{2},\label{eq:yncoeff}\\
    y_2=&
        \frac{a_1(2a_1-c_1-d_1)-a_2}{2v^2}-\frac{(c_1-d_1)^2 -4(c_2+d_2)}{8},\nn
\end{align}
(See Supplement Material \cite{suppmat} for $y_{3,4}$).
Of course, if one is interested in the result in the limit $b\to b_c$, then the required order would be so high that the metric functions have to be known completely.

The rest of the computation, in particular the integrability of the deflection angle are exactly the same as in the SAS case, with the final result
\be
\alpha=\sum_{n=1}^\infty l_n(\beta_s,~\beta_d)\frac{y_n}{b^n}
-2i\pi \label{eq:alpharess}
\ee
where $l_n$ is still given by Eq. \eqref{eq:cndef}.
Ignoring the factor $l_n(\beta_s,~\beta_d)$ which is known to be not in power form of $1/b$, the rest of this deflection angle is automatically arranged in powers of $1/b$. This indeed characterizes the main difference from the formula \eqref{eq:alphares} in the SAS case.

\section{Weak field limit\label{sec:wfl}}
Although the results \eqref{eq:alphares} and \eqref{eq:alpharess} can approximate very large deflection angles, presently and in the near future only deflection angle in the weak field limit, i.e., $b\gg m$, is needed for observations. Indeed, much of the terms in formula
\eqref{eq:alphares} and \eqref{eq:alpharess}, namely the $y_n /b^n$, are already in a series form of the small quantity $m/b$. The weak field limit will then enable us to expand the only left factor $l_n$ also into series of $m/b$ through the expansion of $\beta_{s,d}$ in Eqs. \eqref{eq:betadef} and \eqref{eq:betadefs} and further through expansion of $p(b,1/r_{s,d})$ in Eqs. \eqref{eq:pdef} and $p(1/r_{s,d})$ in \eqref{eq:pdefs}. Because $r_{s,d}$ are also present in $l_n$s, this in turn allows us to see more transparently the effect of finite distance.

For the SAS spacetimes, we will not expand formula \eqref{eq:pdef} for general metrics because of the excessive length of the result, although it is algebraically simple. Rather, we only do a large $b$ expansion for the function \eqref{eq:pdefkn} of the KN spacetime. Further substituting into \eqref{eq:betadef}, \eqref{eq:cndef} and then \eqref{eq:alpharesf}, one finds the KN deflection angle to the second order of $m/b$
\bea
\alpha&=&\sum_{i=s,d}\lsb \frac{m}{b}\left(1+\frac1{v^2}\right)c_{i0}
+\frac{m^2}{b^2} \left\{\frac{s_{i0}^3}{c_{i0}}\left(\frac{1}{v^2} +\frac1{v^4}\right)\right.\right.\nn\\
&&-\frac{2s \hat{a} c_{i0}}{v}-\frac{1}{4} \left(\frac{\pi}{2} -\beta_i+s_{i0}c_{i0}\right)\nn\\
&&\left.\left.\times \left[\hat{q}^2\left(1+\frac2{v^2}\right)-3 \left(1+\frac4{v^2}\right)\right]\right\}\rsb +\mathcal{O}\lb\frac{m}{b}\rb^3\label{eq:alphalbsas}
\eea
where $\beta_{i0}=\arcsin (b/r_i)$ are the zero order apparent angles, and $s_{i0}=\sin\beta_{i0}$ and $c_{i0}=\cos\beta_{i0}$ ($i=s,d$) (see Supplement Material \cite{suppmat} for higher orders).
The Kerr case with $\hat{q}=0$ of this result appeared in Ref. \cite{Li:2019qyb} and partially so in Ref. \cite{Ono:2019hkw} for lightray.

For SSS spacetimes, it was known that when $r_{s,d}$ are infinite and $b\gg m$, the whole deflection angle \eqref{eq:alpharess} to the $n$-th order of $1/b$ can be completely determined by the asymptotic expansion of the metric functions \eqref{eq:sssasyexp} to the $n$-th order \cite{Jia:2020dap}.
Now in the finite $r_{s,d}$ case, one can show without much difficulty that the factor $y_n$ and the large $b$ expansion of factor $l_n$ in Eq. \eqref{eq:alpharess} can both be determined by the expansion \eqref{eq:sssasyexp} to the $n$-th order.
This implies that in the large $b$ limit, the deflection angle \eqref{eq:alpharess}, regardless whether the source and detector are at finite or infinite distances, can be completely determined to the $(m/b)^n$ order by the expansion \eqref{eq:sssasyexp} to the $n$-th order. In particular, for asymptotically flat spacetime, after setting $a_1=-2m$ and $c_n=0$ in Eq. \eqref{eq:sssasyexp} without losing any generality \cite{Weinberg:1972kfs}, the deflection angle to the order $(m/b)^2$ can be shown to equal
\bea
\alpha&=&\sum_{i=s,d}\lsb \frac {m}b \left(\frac{d_1}{2m}+\frac {1}{v^2}\right)c_{i0}
+\frac{m^2}{b^2}\left\{\frac{s_{i0}^3}{2c_{i0}}\left(\frac{d_1}{v^2m} +\frac2{v^4}\right)\right.\right.\nn\\
&&\left.\left.+\left(\frac{\pi}{2} -\beta_i+ s_{i0}c_{i0}\right) \left[\frac{2}{v^2}+\frac{d_1}{2v^2m}-\frac{d_1^2}{16m^2}\right.\right.\right.\nn\\
&&\left.\left.\left.-
\frac{a_2}{2v^2m^2}+\frac{d_2}{4m^2}\right]\right\}
\rsb +\mathcal{O}\lb\frac{m}{b}\rb^3
\label{eq:alphalbsss}
\eea
where again
$\beta_{i0}=\arcsin(b/r_i)$, $s_{i0}=\sin\beta_{i0}$ and $c_{i0}=\cos\beta_{i0}$ ($i=s,d$) (again see Supplement Material \cite{suppmat} for higher orders).

\section{Discussions}
We now discuss the implication of the result and possible extensions of the method developed here. In Ref. \cite{Jia:2020dap}, it was shown that in the equatorial plane of the SAS spacetimes or for any geodesics in the SSS spacetimes, the deflection angle to the first $n$ orders of $m/b$ with infinite $r_{s,d}$ can be determined for metrics that is only asymptotically known to the first $n$ order. In this paper we have shown that this is also the case for the deflection angles with finite source and detector distances, as suggested by formulas \eqref{eq:alphalbsas} and \eqref{eq:alphalbsss}.

The deflection angles \eqref{eq:alphalbsas} and \eqref{eq:alphalbsss} are obtained by only expanding in the large $b/m$ limit. In practice, there are two more limits one can take for these formulas, the ultra-relativistic limit $v\to c$ for particles such as supernova neutrinos and GWs and the approximation $r_s,r_d\gg b$ for practical GL observations. These limits are quite simple and one can directly take them if interested.

As for the equatorial motion in asymptotically non-flat SAS and SSS spacetimes, a brief analysis of our method shows that starting from definition \eqref{asaiint}, an effective deflection angle can still be computed using this procedure, resulting in a formula in terms of $L,~E$. Although $b$ in Eq. \eqref{eq:ledef} can only be interpreted as an impact parameter in the asymptotically flat spacetimes \cite{Butcher:2016yrs}, one can still express the deflection angle using an effective $b$ and a locally defined velocity in the weak curvature limit (e.g., $\Lambda\to 0$ in the (A)dS spacetimes) when $r_{s,d}$ is within the outer horizon (if any).

With the method developed here, the next step is to apply it to the deflection angle and GL in various SAS or SSS spacetimes. It would be interesting to see how the high order parameters of the spacetimes and parameters of the signals (velocity, mass, mass order etc.) can affect the observables in GLs. For the former, including $a_n,~b_n~(n\geq2)$ and $c_n$ in Eq.  \eqref{eq:sssasyexp} and similar expansion coefficient in SAS metrics, we see that they only appear at or above the $(m/b)^2$ order in Eq. \eqref{eq:alphalbsss}. A simple estimation can show that when $a_n/m^n,~b_n/m^n~(n\geq2)$ are not very large, the $(m/b)^2$ or higher order contribution to $\alpha$ can only cause a change of the GL observables---image apparent angles and magnification---far beyond the capability of current or near future observatories. This immediately implies that for arbitrary SSS or SAS spacetimes, effects of high order parameters of moderate size will not be recognizable soon in GL observations through these observables.

Lastly, one might notice that the essential steps in method developed here is the change of variable \eqref{eq:udef} and the expansion \eqref{eq:yexp}. These steps
are applicable to calculations involving other geodesic integrals such as the time delay in the GLs. Corresponding work in this direction will be published elsewhere.

\acknowledgments
We thank Dr. Nan Yang and Mr. Haotian Liu for helpful discussions. This work is supported by the NNSF China 11504276 and MOST China 2014GB109004.

\appendix

\section{Deflection angle in KN spacetime \label{sec:appkn}}

Although deflection angles of very high order are rarely needed in current and near future GL observations, here we present the result in the equatorial plane of the KN spacetime to the 9-th order just to illustrate the power of the method developed in this work. In next appendix, results in general SSS spacetimes will be given.

In the equatorial plane of the KN spacetime, for a signal originating from a source located at $r_s$ and received by a detector at $r_d$, the deflection angle has been obtained to order $(m/b)^2$ in Eq. \eqref{eq:alphasasinz}
with the coefficients given in Eq. \eqref{eq:zsmall}.
Here we present the coefficients in Eq. \eqref{eq:alphasasinz} from $z_3$ to $z_9$
\begin{widetext}
\begin{subequations} \label{eq:zhigh}
\begin{align}
    z_3=
        &
        \frac{l_3}{2} \left[5+45 \frac1{v^2}+
        \frac{15}{v^4}-\frac1{v^6}-3 \hat{q}^2 \left(1+
        \frac6{v^2}+\frac1{v^4}\right)\right] +\frac{2l_2s\hat{a}}v \left[6+ \frac4{v^2}-\hat{q}^2\right] + \frac{3l_3}{2} \hat{a}^2 \left[1+\frac1{v^2}\right],\\
z_4=
        &
        \frac{l_4}{8} \left[35 \left(1+\frac{16}{v^2}+
        \frac{16}{v^4}\right)-30 \hat{q}^2 \left(1+ \frac{12}{v^2}+\frac8{v^4}\right)+\hat{q}^4 \left(3+
        \frac{24}{v^2}+\frac8{v^4}\right)\right] \nn\\
        & +\frac{9 l_3 s\hat{a}}v \left[5+\frac{10}{v^2}+\frac1{v^4}-2
        \hat{q}^2 \left(1+\frac1{v^2}\right)\right]
        +\frac{\hat{a}^2 }{2} \left\{\left[15 l_4+(3 l_2+5 l_4) \frac{8}{v^2}+\frac{8 l_4}{v^4}\right]-l_4\left(3 +\frac{4}{v^2}\right) \hat{q}^2\right\}+\frac{3 l_3s\hat{a}^3 }v ,\\
    z_5=
        &
        \frac{l_5}{8}\left\{3
        \left(21+\frac{525 }{v^2}+\frac{1050 }{v^4}+\frac{210 }{v^6}-
        \frac{15}{v^8}+\frac1{v^{10}}\right)-10 \hat{q}^2 \left(7+
        \frac{140}{v^2}+\frac{210}{v^4}+\frac{28}{v^6}-\frac1{v^8}\right)\right.\nonumber\\
        &
        \left.+15 \hat{q}^4 \left(1+
        \frac{15}{v^2}+ \frac{15}{v^4}+\frac1{v^6}\right)\right\}
        +\frac{2l_4s\hat{a}}v \left\{14\left(5+\frac{20}{v^2}- \frac8{v^4}\right)+3 \hat{q}^2\left(15+\frac{40}{v^2}+\frac8{v^4}\right)
        +\hat{q}^4 \left(3+\frac4{v^2}\right)\right\}\nonumber\\
        &
        +\frac{3}{4} \hat{a}^2 \left\{\left[35 l_5+(160 l_3+175 l_5) \frac{1}{v^2}+(96 l_3+105 l_5)
        \frac{1}{v^4}+\frac{5 l_5 }{v^6}\right]-\hat{q}^2\left[15
        l_5+(32 l_3+50 l_5) \frac{1}{v^2}+\frac{15 l_5}{v^4}\right]\right\}\nonumber\\
        &
        +\frac{4 l_4s\hat{a}^3}v \left(10+ \frac{8}{v^2}-\hat{q}^2\right)
        +\frac{15}{8} l_5 \hat{a}^4\left(1+\frac1{v^2}\right),\\
    z_6=
        &
        \frac{ l_6}{16}\left[
            231 \left(1+\frac{36}{v^2}+\frac{120}{v^4}+\frac{64}{v^6}\right)
            -315 \hat{q}^2 \left(1+\frac{30}{v^2}+\frac{80}{v^4}+\frac{32}{v^6}\right)\right.\nn\\
            &
            \left.+21 \hat{q}^4 \left(5+\frac{120}{v^2}+\frac{240}{v^4}+\frac{64}{v^6}\right)
            -\hat{q}^6 \left(5+\frac{90}{v^2}+\frac{120}{v^4}+\frac{16}{v^6}\right)\right]\nonumber\\
        &
        + \frac{25 l_5s\hat{a}}{4v}\left[63+\frac{420}{v^2}+\frac{378}{v^4}+ \frac{36}{v^6}-\frac1{v^8}+3 \hat{q}^4
        \left(3+\frac{10}{v^2}+\frac3{v^4}\right)-8 \hat{q}^2 \left(7+
        \frac{35}{v^2}+\frac{21}{v^4}+\frac1{v^6}\right)\right]\nonumber\\
        &
        +\frac{1}{4}  \hat{a}^2 \left\{\left[315 l_6+(10 l_4+9 l_6) \frac{280 }{v^2}+(40 l_4+27 l_6)\frac{112 }{v^4}+(10 l_4+9 l_6)\frac{64 }{v^6}\right]\right.\nonumber\\
        &
        \left.-\hat{q}^2\left[210 l_6+ (20 l_4+21
        l_6) \frac{60}{v^2}+(20 l_4+21 l_6) \frac{48 }{v^4}+\frac{96 l_6}{v^6}\right] +\hat{q}^4\left[15 l_6+(2l_4+3l_6) \frac{20}{v^2}+\frac{24 l_6 }{v^4}\right]
        \right\}\nonumber\\
        &
        +\frac{5s\hat{a}^3}{2v}
        \left\{\left[105 l_5+(32 l_3+210 l_5) \frac{1}{v^2}+\frac{45 l_5}{v^4}\right]-30 l_5 \hat{q}^2\left(1+\frac{1}{v^2}\right)\right\}\nonumber\\
        &
        +\frac{\hat{a}^4 }{2}\left\{\left[35 l_6+(20l_4+21l_6) \frac{4}{v^2}+   \frac{24 l_6}{v^4}\right]-\left(5 l_6+\frac{6 l_6}{v^2}\right) \hat{q}^2\right\}
        +\frac{15l_5s\hat{a}^5 }{4v},\\
    z_7=
        &
        \frac{l_7}{16}\left[\left(429+\frac{21021 }{v^2}+\frac{105105 }{v^4}+\frac{105105 }{v^6}+\frac{15015 }{v^8}-\frac{1001 }{v^{10}}+\frac{91}{v^{12}}-\frac{5 }{v^{14}}\right)\right.\nonumber\\
        &
        -21 \hat{q}^2 \left(33+\frac{1386 }{v^2}+\frac{5775 }{v^4}+\frac{4620 }{v^6}+\frac{495}{v^8}-\frac{22 }{v^{10}}+\frac{1}{v^{12}}\right)
        +35 \hat{q}^4\left(9+\frac{315 }{v^2}+\frac{1050 }{v^4}+\frac{630}{v^6} +\frac{45 }{v^8}-\frac{1}{v^{10}}\right)\nn\\
        &
        \left. -35 \hat{q}^6 \left(1+\frac{28 }{v^2}+\frac{70 }{v^4}+\frac{28}{v^6}+\frac{1}{v^8}\right) \right]
        +\frac{9l_6s\hat{a}}{4v}\left[66 \left(7+\frac{70 }{v^2}+\frac{112 }{v^4}+\frac{32 }{v^6}\right)\right.\nn\\
        &
        \left.-15 \hat{q}^2\left(35+\frac{280 }{v^2}+\frac{336 }{v^4}+\frac{64}{v^6}\right) +\hat{q}^4\left(140+\frac{840 }{v^2}+\frac{672 }{v^4}+\frac{64 }{v^6}\right)
        -\hat{q}^6\left(5+\frac{20 }{v^2}+\frac{8 }{v^4}\right) \right]\nonumber\\
        &
        +\frac{5\hat{a}^2}{16}\left\{\left[693 l_7+\frac{105 }{v^2}\left(96 l_5+77 l_7\right)+\frac{210 }{v^4}\left(144
        l_5+77 l_7\right)+\frac{90 }{v^6}\left(144 l_5+77
        l_7\right)+\frac{5 }{v^8}\left(96 l_5+77 l_7\right)-\frac{7l_7}{v^{10}}\right]\right.\nonumber\\
        &
        \left.-10\hat{q}^2\left[63 l_7+\frac{84 }{v^2}\left(8 l_5+7 l_7\right)+\frac{42}{v^4}\left(32l_5+21 l_7\right)+\frac{36 }{v^6}\left(8 l_5+7
        l_7\right)+\frac{7 l_7}{v^8}\right]\right.\nonumber\\
        &
        \left.+15\hat{q}^4\left[7 l_7+\frac{1}{v^2}\left(48 l_5+49 l_7\right)+\frac{1}{v^4}\left(48 l_5+49 l_7\right)+\frac{7
        l_7}{v^6}\right]\right\}\nn\\
        &
        +\frac{s\hat{a}^3}{v}\left\{4\left[315 l_6+\frac{14 }{v^2}\left(20 l_4+81 l_6\right)+\frac{8}{v^4} \left(20 l_4+81 l_6\right)+\frac{48 l_6}{v^6}\right]\right.\nonumber\\
        &
        \left.-6\hat{q}^2\left[105 l_6+\frac{1}{v^2}\left(40 l_4+252 l_6\right)+\frac{72l_6}{v^4}\right]+6l_6\hat{q}^4\left(5+\frac{6}{v^2}\right)\right\}\nonumber\\
        &
        +\frac{15\hat{a}^4}{16}\left\{\left[105 l_7+\frac{1}{v^2}\left(672 l_5+441 l_7\right)+\frac{1}{v^4}\left(480 l_5+315l_7\right)+\frac{35 l_7}{v^6}\right]
        -\hat{q}^2\left(35 l_7+\frac{96 l_5+98 l_7}{v^2}+\frac{35l_7}{v^4}\right)\right\}\nonumber\\
        &
        +\frac{6l_6s\hat{a}^5}{v}\left[\left(14+\frac{12}{v^2}\right)-\hat{q}^2\right]
        +\frac{35l_7\hat{a}^6}{16}\left(1+\frac{1}{v^2}\right),
        \\
    z_8=
        &
        \frac{l_8}{128}\left[1287\left(5+\frac{320}{v^2}+\frac{2240}{v^4}
        +\frac{3584}{v^6}+\frac{1280}{v^8}\right)
        -12012\hat{q}^2\left(1+\frac{56}{v^2}+\frac{336}{v^4}+\frac{448}{v^6}
        +\frac{128}{v^8}\right)\right.\nonumber\\
        &
        +990\hat{q}^4\left(7+\frac{336}{v^2}+\frac{1680}{v^4}+\frac{1792}{v^6}
        +\frac{384}{v^8}\right)
        -180\hat{q}^6\left(7+\frac{280}{v^2}+\frac{1120}{v^4}+\frac{896}{v^6}
        +\frac{128}{v^8}\right)\nonumber\\
        &
        \left.+\hat{q}^8\left(35+\frac{1120}{v^2}+\frac{3360}{v^4}+\frac{1792}{v^6}
        +\frac{128}{v^8}\right)\right]
        +\frac{49l_7s\hat{a}}{8v}\left[\left(429+\frac{6006}{v^2}+\frac{15015}{v^4}
        +\frac{8580}{v^6}\right.\right.\nn\\
        &
        \left.+\frac{715}{v^8}-\frac{26}{v^{10}}+\frac{1}{v^{12}}\right)
        -\hat{q}^2\left(594+\frac{6930}{v^2}+\frac{13860}{v^4}+\frac{5940}{v^6}
        +\frac{330}{v^8}-\frac{6}{v^{10}}\right)\nn\\
        &
        \left.+25\hat{q}^4\left(9+\frac{84}{v^2}+\frac{126}{v^4}+\frac{36}{v^6}
        +\frac{1}{v^8}\right)
        -20\hat{q}^6\left(1+\frac{7}{v^2}+\frac{7}{v^4}+\frac{1}{v^6}\right)\right]\nonumber\\
        &
        +\frac{\hat{a}^2}{16}\left\{33\left[273 l_8+\frac{168 }{v^2}\left(35 l_6+26 l_8\right)+\frac{1008 }{v^4}\left(28l_6+13 l_8\right)+\frac{384 }{v^6}\left(63 l_6+26l_8\right)+\frac{128 }{v^8}\left(28 l_6+13l_8\right)\right]\nonumber\right.\\
        &
        \left.-15\hat{q}^2\left[693 l_8+\frac{840 }{v^2}\left(14 l_6+11 l_8\right)+\frac{2016}{v^4} \left(21 l_6+11 l_8\right)+\frac{1152}{v^6}\left(21 l_6+11 l_8\right)+\frac{128 }{v^8}\left(14 l_6+11 l_8\right)\right]\right.\nonumber\\
        &
        \left.+9\hat{q}^4\left[315 l_8+\frac{560 }{v^2}\left(7 l_6+6 l_8\right)+\frac{672 }{v^4}\left(14 l_6+9 l_8\right)+\frac{384}{v^6}\left(7 l_6+6 l_8\right)+\frac{128 l_8}{v^8}\right]\right.\nonumber\\
        &
        \left.-3\hat{q}^6\left[35 l_8+\frac{280 }{v^2}\left(l_6+l_8\right)+\frac{336 }{v^4} \left(l_6+l_8\right)+\frac{64 l_8}{v^6}\right]\right\}\nonumber\\
        &
        +\frac{35s\hat{a}^3}{8v}\left\{\left[1155 l_7+\frac{84 }{v^2}\left(24 l_5+77 l_7\right)+\frac{90}{v^4}\left(32 l_5+77 l_7\right)+\frac{20 }{v^6}\left(24 l_5+77 l_7\right)+\frac{35 l_7}{v^8}\right]\right.\nonumber\\
        &
        \left.-8\hat{q}^2\left(105 l_7+\frac{112 l_5+441 l_7}{v^2}+\frac{80 l_5+315 l_7}{v^4}+\frac{35 l_7}{v^6}\right)+\hat{q}^4\left(105 l_7+\frac{1}{v^2}\left(48 l_5+294 l_7\right)+\frac{105 l_7}{v^4}\right)\right\}\nonumber\\
        &
        +\hat{a}^4\left\{\left[\frac{3465 l_8}{8}+\frac{252}{v^2}\left(21 l_6+11 l_8\right)+\frac{1}{v^4}\left(560 l_4+9072 l_6+3564 l_8\right)+\frac{96 }{v^6} \left(21 l_6+11 l_8\right)+\frac{48 l_8}{v^8}\right]\right.\nonumber\\
        &
        \left.-\frac{9\hat{q}^2}{4}\left[105 l_8+\frac{1}{v^2}\left(784 l_6+504 l_8\right)+\frac{1}{v^4}\left(672 l_6+432 l_8\right)+\frac{64 l_8}{v^6}\right]+\hat{q}^4\left[\frac{105 l_8}{8}+\frac{42 }{v^2}\left(l_6+l_8\right)+\frac{18l_8}{v^4}\right]\right\}\nonumber\\
        &
        +\frac{105s\hat{a}^5}{8v}\left\{\left[63 l_7+\frac{1}{v^2}\left(32 l_5+126 l_7\right)+\frac{35 l_7}{v^4}\right]-14l_7\hat{q}^2\left(1+\frac1{v^2}\right)\right\}\nonumber\\
        &
        +\hat{a}^6\left\{\frac32\left(21 l_8+\frac{8}{v^2}\left(7 l_6+6 l_8\right)+\frac{16 l_8}{v^4}\right)-\frac{l_8 \hat{q}^2}{2} \left(7+\frac{8}{v^2}\right)\right\}+\frac{35 l_7 s\hat{a}^7}{8 v},
        \\
    z_9=
        &
        \frac{l_9}{128}\left[\left(12155+\frac{984555}{v^2}+\frac{9189180}{v^4}+\frac{21441420}{v^6}
        +\frac{13783770}{v^8}+\frac{1531530}{v^{10}}-\frac{92820}{v^{12}}+\frac{ 9180}{v^{14}}-\frac{765}{v^{16}}+\frac{35}{v^{18}}\right)\right.\nonumber\\
        &
        -180\hat{q}^2\left(143+\frac{10296}{v^2}+\frac{84084}{v^4}+\frac{168168}{v^6}
        +\frac{90090 }{v^8}+\frac{8008}{v^{10}}-\frac{364}{v^{12}}
        +\frac{24}{v^{14}}-\frac{1}{v^{16}}\right)\nonumber\\
        &
        +126\hat{q}^4\left(143+\frac{9009}{v^2}+\frac{63063}{v^4}+\frac{105105}{v^6}+\frac{45045}
   {v^8}+\frac{3003}{v^{10}}-\frac{91}{v^{12}}+\frac{3}{v^{14}}\right)\nonumber\\
        &
        -\left.420\hat{q}^6\left(11+\frac{594}{v^2}+\frac{3465}{v^4}+\frac{4620}{v^6}+\frac{1485}{v^8}+
        \frac{66}{v^{10}}-\frac{1}{v^{12}}\right)
        +315\hat{q}^8\left(1+\frac{45}{v^2}+\frac{210}{v^4}+\frac{210}{v^6}+\frac{45}{v^8}+
        \frac{1}{v^{10}}\right)\right]\nonumber\\
        &
        +\frac{l_8s\hat{a}}{2v}\left[286\left(45+\frac{840}{v^2}+\frac{3024}{v^4}
        +\frac{2880}{v^6}+\frac{640}{v^8}\right)
        -1001\hat{q}^2\left(21+\frac{336}{v^2}+\frac{1008}{v^4}+\frac{768}{v^6}+\frac{128}{v^8}\right)\right.\nonumber\\
        &
        +165\hat{q}^4\left(63+\frac{840}{v^2}+\frac{2016}{v^4}+\frac{1152}{v^6}+\frac{128}{v^8}\right)
        -5\hat{q}^6\left(315+\frac{3360}{v^2}+\frac{6048}{v^4}+\frac{2304}{v^6}+\frac{128}{v^8}\right)\nonumber\\
        &
        \left.+\hat{q}^8\left(35+\frac{280}{v^2}+\frac{336}{v^4}+\frac{64}{v^6}\right)\right\}
        +\frac{7\hat{a}^2}{32}\left\{\left[6435 l_9+\frac{3003}{v^2}\left(64 l_7+45 l_9\right)+\frac{21021}{v^4}\left(64 l_7+27 l_9\right)\right.\right.\nonumber\\
        &
        \left.+\frac{15015 }{v^6}\left(128 l_7+45 l_9\right)+\frac{5005 }{v^8}\left(128 l_7+45 l_9\right)+\frac{455 }{v^{10}}\left(64 l_7+27 l_9\right)-\frac{7 }{v^{12}}\left(64 l_7+45 l_9\right)+\frac{9 l_9}{v^{14}}\right]\nonumber\\
        &
        -21\hat{q}^2\left[429 l_9+\frac{66}{v^2} \left(160 l_7+117 l_9\right)+\frac{231 }{v^4}\left(256 l_7+117 l_9\right)+\frac{1980 }{v^6}\left(32 l_7+13 l_9\right)\right.\nonumber\\
        &
        \left.+\frac{55 }{v^8}\left(256 l_7+117 l_9\right)+\frac{1}{v^{10}}\left(320 l_7+234 l_9\right)-\frac{3 l_9}{v^{12}}\right]\nonumber\\
        &
        +35\hat{q}^4\left[99 l_9+\frac{15 \left(128 l_7+99 l_9\right)}{v^2}+\frac{126 \left(64 l_7+33 l_9\right)}{v^4}+\frac{90 \left(64 l_7+33 l_9\right)}{v^6}+\frac{640 l_7+495 l_9}{v^8}+\frac{9 l_9}{v^{10}}\right]\nn\\
        &
        \left.-7\hat{q}^6\left[45 l_9+\frac{1}{v^2}\left(640 l_7+540 l_9\right)+\frac{1}{v^4}\left(1792 l_7+1134 l_9\right)+\frac{1}{v^6}\left(640 l_7+540 l_9\right)+\frac{45 l_9}{v^8}\right]\right\}\nonumber\\
        &
        +\frac{s\hat{a}^3}{v}\left\{\left[18018 l_8+\frac{3696}{v^2} \left(14 l_6+39 l_8\right)+\frac{19008 }{v^4}\left(7 l_6+13 l_8\right)+\frac{8448 }{v^6}\left(7 l_6+13 l_8\right)+\frac{256}{v^8} \left(14 l_6+39 l_8\right)\right]\right.\nonumber\\
        &
        -15\hat{q}^2\left[1155 l_8+\frac{336 }{v^2}\left(7 l_6+22 l_8\right)+\frac{288}{v^4} \left(14 l_6+33 l_8\right)+\frac{128 }{v^6}\left(7 l_6+22 l_8\right)+\frac{128 l_8}{v^8}\right]\nonumber\\
        &
        \left.+\hat{q}^4\left[3780 l_8+\frac{672 }{v^2}\left(7 l_6+27 l_8\right)+\frac{576 }{v^4}\left(7 l_6+27 l_8\right)+\frac{2304 l_8}{v^6}\right]
        -\hat{q}^6\left[105 l_8+\frac{56 \left(l_6+6 l_8\right)}{v^2}+\frac{144 l_8}{v^4}\right]\right\}\nonumber\\
        &
        +\frac{35\hat{a}^4}{64}\left\{\left[3003 l_9+\frac{231 \left(256 l_7+117 l_9\right)}{v^2}+\frac{2304 \left(8 l_5+77 l_7\right)+54054 l_9}{v^4}+\frac{1280 \left(8 l_5+77 l_7\right)+30030 l_9}{v^6}\right.\right.\nonumber\\
        &
        \left.+\frac{8960 l_7+4095 l_9}{v^8}+\frac{63 l_9}{v^{10}}\right]
        -2\hat{q}^2\left[1155 l_9+\frac{252 \left(64 l_7+33 l_9\right)}{v^2}+\frac{512 \left(4 l_5+63 l_7\right)+12474 l_9}{v^4}\right.\nonumber\\
        &
        \left.\left.+\frac{140 \left(64 l_7+33 l_9\right)}{v^6}+\frac{315 l_9}{v^8}\right]+21\hat{q}^4\left(15 l_9+\frac{128 l_7+81 l_9}{v^2}+\frac{128 l_7+81 l_9}{v^4}+\frac{15 l_9}{v^6}\right)\right\}\nn\\
        &
        +\frac{12s\hat{a}^5}{v}\left[\left(462 l_8+\frac{672 l_6+1584 l_8}{v^2}+\frac{448 l_6+1056 l_8}{v^4}+\frac{128 l_8}{v^6}\right)-\hat{q}^2\left(189 l_8+\frac{112 l_6+432 l_8}{v^2}+\frac{144 l_8}{v^4}\right)\right.\nonumber\\
        &
        \left.
        +l_8 \hat \hat{q}^4\left(7+\frac{8}{v^2}\right)\right]
        +\frac{105\hat{a}^6}{32}\left[\left(77 l_9+\frac{576 l_7+297 l_9}{v^2}+\frac{448 l_7+231 l_9}{v^4}+\frac{35 l_9}{v^6}\right)\right.\nonumber\\
        &
        \left.-\hat{q}^2\left(21 l_9+\frac{64 l_7+54 l_9}{v^2}+\frac{21 l_9}{v^4}\right)\right]
        +\frac{8l_8s\hat{a}^7}{v}\left[\left(18+\frac{16}{v^2}\right)-\hat{q}^2\right]
        +\frac{315 l_9\hat{a}^8}{128}\left(1+\frac1{v^2}\right).
\end{align}
\end{subequations}
\end{widetext}

The large $b$ limit of the deflection angle \eqref{eq:alphasasinz} to the second order was obtained in Eq. \eqref{eq:alphalbsas}. Here we extend this result to third order
\begin{widetext}
\begin{align}
\alpha=&\sum_{i=s,d}\lsb \frac{m}{b}\left(1+\frac1{v^2}\right)c_{i0}
        +\left(\frac{m}{b}\right)^2\left\{\frac{s_{i0}^3}{c_{i0}}\left(\frac{1}{v^2} +\frac1{v^4}\right)+\frac{2 \hat{a}s c_{i0}}{v}-\frac{1}{8} (\pi -2\beta_{i0}+2s_{i0}c_{i0}) \left[\hat{q}^2\left(1+\frac2{v^2}\right)-3 \left(1+\frac4{v^2}\right)\right]\right\}\right.\nonumber\\
        &
        +\left(\frac{m}{b}\right)^3\left\{\hat{a}^2\frac{ \left(1+c_{i0}^2\right)}{2c_{i0}}\left(1+\frac1 {v^2}\right)
        +\frac{\hat{a}s}{v}\left[-\frac{\hat{q}^2}{2}\left(\pi+2s_{i0}c_{i0}-2\beta_{i0}\right)
        +\left(\pi +\frac{2 s_{i0}}{c_{i0}}-2\beta_{i0}\right) \left(3+\frac2{v^2}\right)-\frac{4s_{i0}^3}{c_{i0}}\right]\right.\nonumber\\
        &
        -\hat{q}^2\left[\frac{1+c_{i0}^2}{2c_{i0}}\left(1+\frac6 {v^2}+\frac1{v^4}\right)-\frac{s_{i0}^4}{2c_{i0}}\left(1+\frac4 {v^2}-\frac2{v^4}\right)\right]+\frac{-2+3 s_{i0}^2-12 s_{i0}^4+8 s_{i0}^6}{6 c_{i0}^3 v^6}+\frac{10-15 s_{i0}^2+12 s_{i0}^4-8 s_{i0}^6}{2 c_{i0}^3 v^4}\nonumber\\
        &\left.
        \left.+\frac{30-15 s_{i0}^2-8 s_{i0}^4}{2 c_{i0} v^2}+\frac{5}{6} c_{i0}
   \left(2+s_{i0}^2\right)\right\}\rsb.
\end{align}
\end{widetext}

\section{Deflection angle in general SSS spacetimes \label{sec:appsss}}

The general metric we used for the SSS spacetime is given in Eq. \eqref{sss1}.
Assuming the asymptotic expansions of the metric functions take the form of Eq. \eqref{eq:sssasyexp}, then the deflection angle in this spacetime was given in Eq. \eqref{eq:alpharess}
with coefficients $y_1$ and $y_2$ given in Eq. \eqref{eq:yncoeff}.
Here we present the $y_n$ to the 4th order
\begin{widetext}
\begin{subequations}
\begin{align}
    y_3=&
        \left(\frac{1}{16 v^6}-\frac{3}{4 v^4}-\frac{3}{2 v^2}\right) a_1^3+\left(\frac{3}{4 v^4}+\frac{3}{v^2}\right) a_1 a_2-\frac{3 a_3}{2 v^2}+\left[\left(\frac{3}{16 v^4}+\frac{3}{4 v^2}\right) a_1^2-\frac{3 a_2}{4 v^2}\right]d_1+\frac{3 a_1 d_1^2}{16 v^2}+\frac{d_1^3}{16}\nn\\
        &
        -\left[\frac{3 a_1}{4 v^2}+\frac{d_1}{4}\right]d_2+\frac{d_3}{2}
        +\left[\left(\frac{3}{8 v^4}+\frac{3}{2 v^2}\right) a_1^2-\frac{3 a_2}{2 v^2}-\frac{3 a_1 d_1}{4 v^2}-\frac{1}{8} d_1^2+\frac{d_2}{2}\right]c_1
        -\left[\frac{3 a_1}{2 v^2}+\frac{d_1}{2}\right]c_2+c_3\\
    y_4=
        &
        \left(\frac{3}{v^4}+\frac{2}{v^2}\right) a_1^4+\left(-\frac{6}{v^4}-\frac{6}{v^2}\right) a_1^2 a_2+\left(\frac{1}{v^4}+\frac{2}{v^2}\right) a_2^2+\left(\frac{2}{v^4}+\frac{4}{v^2}\right) a_1 a_3-\frac{2 a_4}{v^2}+\left[-\left(\frac{1}{v^4}+\frac{1}{v^2}\right) a_1^3 \right.\nn\\
        &
        \left.+\left(\frac{1}{v^4}+\frac{2}{v^2}\right) a_1 a_2-\frac{a_3}{v^2}\right]d_1+\left[-\left(\frac{1}{8 v^4}+\frac{1}{4 v^2}\right) a_1^2+\frac{a_2}{4 v^2}\right] d_1^2-\frac{a_1 d_1^3}{8 v^2}-\frac{5 d_1^4}{128}+\left[\left(\frac{1}{2 v^4}+\frac{1}{v^2}\right) a_1^2-\frac{a_2}{v^2}\right.\nn\\
        &
        \left.+\frac{a_1 d_1}{2 v^2}+\frac{3}{16} d_1^2\right] d_2-\frac{d_2^2}{8}-\left[\frac{a_1}{v^2}+\frac{d_1}{4}\right] d_3+\frac{d_4}{2}+\left[-\left(\frac{3}{v^4}+\frac{3}{v^2}\right) a_1^3+\left(\frac{3}{v^4}+\frac{6}{v^2}\right) a_1 a_2-\frac{3 a_3}{v^2}\right.\nn\\
        &
        \left.+\left(\frac{3}{4 v^4}+\frac{3}{2 v^2}\right) a_1^2 d_1-\frac{3 a_2 d_1}{2 v^2}+\frac{3 a_1 d_1^2}{8 v^2}+\frac{3}{32} d_1^3-\frac{3 a_1 d_2}{2 v^2}-\frac{3}{8} d_1 d_2+\frac{3 d_3}{4}\right]c_1+\left[\left(\frac{3}{8 v^4}+\frac{3}{4 v^2}\right) a_1^2\right.\nn\\
        &
        \left.-\frac{3 a_2}{4 v^2}-\frac{3 a_1 d_1}{8 v^2}-\frac{3}{64} d_1^2+\frac{3}{16} d_2\right]c_1^2+\left[\frac{a_1}{8 v^2}-\frac{d_1}{32}\right]c_1^3+\frac{3 c_1^4}{128}+\left[\left(\frac{3}{2 v^4}+\frac{3}{v^2}\right) a_1^2-\frac{3 a_2}{v^2}-\frac{3 a_1 d_1}{2 v^2}\right.\nn\\
        &
        \left.-\frac{3}{16} d_1^2+\frac{3 d_2}{4}-\frac{3 a_1 c_1}{2 v^2}+\frac{3}{8} d_1 c_1-\frac{3}{16} c_1^2\right]c_2+\frac{3 c_2^2}{8}+\left[-\frac{3 a_1}{v^2}+\frac{3 d_1}{4}+\frac{3 c_1}{4}\right]c_3+\frac{3 c_4}{2}.
\end{align}\label{eq:eqynsssgv}
\end{subequations}
\end{widetext}

In the expansion \eqref{eq:sssasyexp}, one can always set $a_1=-2m$ where $m$ is the ADM mass of the spacetime, and it is often the case that the metric function $C(r)$ is scaled to $C(r)=r^2$ \cite{Weinberg:1972kfs}. With this simplification, the functions $y_n$ in Eq. \eqref{eq:yncoeff} and \eqref{eq:eqynsssgv} become
\begin{widetext}
\begin{subequations}
\begin{align}
    y_1=&m\lb
        \frac{1}{v^2}+\frac{\hat{d}_1}{2}\rb\\
    y_2=&m^2\lb
        \frac{4}{v^2}-\frac{\hat{a}_2}{v^2}+\frac{\hat{d}_1}{v^2}-\frac{
   \hat{d}_1^2}{8}+\frac{\hat{d}_2}{2}\rb\\
    y_3=&m^3\lsb
        \frac{12}{v^2}+\frac{6}{v^4}-\frac{1}{2 v^6}-\left(\frac{3}{2 v^4}+\frac{6}{v^2}\right) \hat{a}_2-\frac{3 \hat{a}_3}{2 v^2}+\left(\frac{3}{4 v^4}+\frac{3}{v^2}\right) \hat{d}_1-\frac{3 \hat{a}_2 \hat{d}_1}{4 v^2}-\frac{3 \hat{d}_1^2}{8 v^2}+\frac{\hat{d}_1^3}{16}+\frac{3 \hat{d}_2}{2 v^2}-\frac{\hat{d}_1 \hat{d}_2}{4} +\frac{\hat{d}_3}{2}\rsb\\
    y_4=&m^4\lsb
        \frac{48}{v^4}+\frac{32}{v^2}-\left(\frac{24}{v^4}+\frac{24}{v^ 2}\right) \hat{a}_2+\left(\frac{1}{v^4}+\frac{2}{v^2}\right) \hat{a}_2^2+\left(-\frac{4}{v^4}-\frac{8}{v^2}\right) \hat{a}_3-\frac{2 \hat{a}_4}{v^2}+\left(\frac{8}{v^4}+\frac{8}{v^2}\right) \hat{d}_1\right.\nn\\
        &
        -\left(\frac{2}{v^4}+\frac{4}{v^2}\right) \hat{a}_2 \hat{d}_1-\frac{\hat{a}_3 \hat{d}_1}{v^2}+\left(-\frac{1}{2 v^4}-\frac{1}{v^2}\right) \hat{d}_1^2+\frac{\hat{a}_2 \hat{d}_1^2}{4 v^2}+\frac{\hat{d}_1^3}{4 v^2}-\frac{5 \hat{d}_1^4}{128}+\left(\frac{2}{v^4}+\frac{4}{v^2}\right) \hat{d}_2-\frac{\hat{a}_2 \hat{d}_2}{v^2}-\frac{\hat{d}_1 \hat{d}_2}{v^2}\nn\\
        &\left. +\frac{3}{16} \hat{d}_1^2 \hat{d}_2-\frac{\hat{d}_2^2}{8}+\frac{2 \hat{d}_3}{v^2}
        -\frac{1}{4} \hat{d}_1 \hat{d}_3+\frac{\hat{d}_4}{2}\rsb
\end{align}\label{eq:ynwithmgv}
\end{subequations}
\end{widetext}
where the hatted quantities are the corresponding quantities scaled by powers of $m$, i.e., $\hat{x}_n=x_n/m^n$ for $x\in\{a,~c,~d\}$.

In the large $b/m$ limit, the deflection angle \eqref{eq:alpharess} with the functions $y_n$ in Eq. \eqref{eq:ynwithmgv} has been expanded to the second order of $m/b$ to yield Eq. \eqref{eq:alphalbsss}. Here we extend this expansion to the third order of $m/b$
\begin{widetext}
\begin{align}
    &\alpha=\sum_{i=s,d}\lsb
        \frac mb \left(\frac{\hat{d}_1}2+\frac1{v^2}\right)c_{i0}
        +\frac{m^2}{b^2}\left[\frac{s_{i0}^3}{v^2c_{i0}}\left(\frac{\hat{d}_1}{2}
        +\frac{1}{v^2}\right)
        +\frac{1}{8}\left(\pi +2 s_{i0}c_{i0}-2\beta\right) \left(\frac{8}{v^2}-\frac{2 \hat{a}_2}{v^2}+\frac{2
   \hat{d}_1}{v^2}-\frac{\hat{d}_1^2}{4}+\hat{d}_2 \right)\right]\right.\nonumber\\
        &
        +\frac{m^3}{b^3}\lcb \frac{s_{i0}^4 \left(-4+3 s_{i0}^2+4 v^2 c_{i0}^2-c_{i0}^2v^2 \hat{a}_2\right) \left(2+v^2
   \hat{d}_1\right)}{4 v^6 c_{i0}^3}+\frac{s_{i0}^4 \left(32-8 \hat{a}_2+8 \hat{d}_1-v^2 \hat{d}_1^2+4 v^2
   \hat{d}_2\right)}{8 v^4 c_{i0}}+\frac{c_{i0}\left(2+s_{i0}^2\right)}{48}\times\right.\nn\\
        &\left.
        \left.\left[-\frac{8}{v^6}+\frac{12 }{v^4}\left(8-2 \hat{a}_2+\hat{d}_1\right)-\frac{6}{v^2}\left(-32+4 \hat{a}_3-8 \hat{d}_1+\hat{d}_1^2+2 \hat{a}_2 \left(8+\hat{d}_1\right)-4 \hat{d}_2\right)+\hat{d}_1^3-4 \hat{d}_1 \hat{d}_2+8 \hat{d}_3\right]\rcb\rsb.
\end{align}
\end{widetext}

Occasionally, the large $b/m$ limit of the change of the angular coordinate in Eq. \eqref{eq:idef3s} is desired for the finite distance case. Here we present its result to the order of $m^2/b^2$,
\begin{widetext}
\begin{align}
\Delta\varphi=&\sum_{i=s,d} \lsb \frac{\pi}2-\beta_{i0}+\frac mb\left(\frac{\hat{c}_1 s_{i0}^2+\hat{d}_1 c_{i0}^2}{2 c_{i0}}+\frac{1}{v^2 c_{i0}}\right)+\frac{m^2}{b^2}\left\{\frac{s_{i0}^3}{c_{i0}} \left(\frac{1}{v^4}+\frac{\hat{c}_1+\hat{d}_1}{2 v^2}+\frac{1}{4} \hat{c}_1 \hat{d}_1\right)\right.\right.\nn\\
&+\frac{s_{i0}^3}{2c_{i0}^3}\left[c_{i0}^2\left(\hat{c_2}+\frac4{v^2} -\frac{\hat{a_2}}{v^2}\right)
-\left(1+2c_{i0}^2\right)\left(\frac{\hat{c}_1^2}{4} +\frac1{v^4}\right)-\frac{\hat{c}_1}{v^2}\right]\nn\\
&\left.\left.
+\left(\frac\pi 2-\beta_{i0}+s_{i0}c_{i0}\right)\left[\frac{-\hat{a}_2+\hat{c}_1+\hat{d}_1+4}{2 v^2}+\frac{\left(\hat{c}_2+\hat{d}_2\right)}{4}-\frac{\left(\hat{c}_1-\hat{d}_1\right)^2}{16}\right]\right\} +\mathcal{O}\lb\frac{m^3}{b^3}\rb \rsb.
\end{align}
\end{widetext}

Finally let us mention that we have applied the method developed here to a few particular spacetimes that we studied, namely the Teo wormhole spacetime \cite{Jia:2020dap}, the Bardeen, Hayward, Janis-Newman-Winicour and Einstein-Born-Infeld spacetimes \cite{Duan:2020tsq}, to investigate the effect of finite $r_{s,d}$ in these spacetimes. Results to higher orders than in this appendix have been obtained, and the previous result with $r_{s,d}\to\infty$ can be fully recovered. The corresponding work will be published elsewhere.




%



\begin{thebibliography}{}
\bibitem{Dyson:1920cwa} F.~W.~Dyson, A.~S.~Eddington and C.~Davidson,
  Phil.\ Trans.\ Roy.\ Soc.\ Lond.\ A {\bf 220}, 291 (1920).
  doi:10.1098/rsta.1920.0009

\bibitem{Sharon:2014ija} K.~Sharon and T.~L.~Johnson,
  Astrophys.\ J.\  {\bf 800}, no. 2, L26 (2015)

\bibitem{Peng:2006ew} C.~Y.~Peng, C.~D.~Impey, H.~W.~Rix, C.~S.~Kochanek, C.~R.~Keeton, E.~E.~Falco, J.~Lehar and B.~A.~McLeod,
  Astrophys.\ J.\  {\bf 649}, 616 (2006)
  [astro-ph/0603248].

\bibitem{Bartelmann:1999yn} M.~Bartelmann and P.~Schneider,
  Phys.\ Rept.\  {\bf 340}, 291 (2001)
  [astro-ph/9912508].

\bibitem{Ade:2015zua} P.~A.~R.~Ade {\it et al.} [Planck Collaboration],
  Astron.\ Astrophys.\  {\bf 594}, A15 (2016)

\bibitem{Refregier:2003ct} A.~Refregier,
  Ann.\ Rev.\ Astron.\ Astrophys.\  {\bf 41}, 645 (2003)
  [astro-ph/0307212].

\bibitem{Lewis:2006fu} A.~Lewis and A.~Challinor,
  Phys.\ Rept.\  {\bf 429}, 1 (2006)
  [astro-ph/0601594].

\bibitem{Metcalf:2001ap} R.~B.~Metcalf and P.~Madau,
  Astrophys.\ J.\  {\bf 563}, 9 (2001)
  [astro-ph/0108224].

\bibitem{Hoekstra:2008db} H.~Hoekstra and B.~Jain,
  Ann.\ Rev.\ Nucl.\ Part.\ Sci.\  {\bf 58}, 99 (2008)

\bibitem{Hirata:1987hu} K.~Hirata {\it et al.} [Kamiokande-II Collaboration],
Phys.\ Rev.\ Lett.\ {\bf 58}, 1490 (1987).

\bibitem{Bionta:1987qt} R.~M.~Bionta {\it et al.},
Phys.\ Rev.\ Lett.\ {\bf 58}, 1494 (1987).

\bibitem{IceCube:2018dnn} M.~G.~Aartsen {\it et al.} [IceCube and Fermi-LAT and MAGIC and AGILE and ASAS-SN and HAWC and H.E.S.S. and INTEGRAL and Kanata and Kiso and Kapteyn and Liverpool Telescope and Subaru and Swift NuSTAR and VERITAS and VLA/17B-403 Collaborations],
  Science {\bf 361}, no. 6398, eaat1378 (2018)

\bibitem{IceCube:2018cha} M.~G.~Aartsen {\it et al.} [IceCube Collaboration],
  Science {\bf 361}, no. 6398, 147 (2018)

\bibitem{Abbott:2016blz} B.~P.~Abbott {\it et al.} [LIGO Scientific and Virgo Collaborations],
  Phys.\ Rev.\ Lett.\  {\bf 116}, no. 6, 061102 (2016)

\bibitem{Abbott:2016nmj} B.~P.~Abbott {\it et al.} [LIGO Scientific and Virgo Collaborations],
  Phys.\ Rev.\ Lett.\  {\bf 116}, no. 24, 241103 (2016)
  doi:10.1103/PhysRevLett.116.241103
  [arXiv:1606.04855 [gr-qc]].

\bibitem{Abbott:2017oio} B.~P.~Abbott {\it et al.} [LIGO Scientific and Virgo Collaborations],
  Phys.\ Rev.\ Lett.\  {\bf 119}, no. 14, 141101 (2017)
  doi:10.1103/PhysRevLett.119.141101
  [arXiv:1709.09660 [gr-qc]].

\bibitem{TheLIGOScientific:2017qsa} B.~P.~Abbott {\it et al.} [LIGO Scientific and Virgo Collaborations],
  Phys.\ Rev.\ Lett.\  {\bf 119}, no. 16, 161101 (2017)
  doi:10.1103/PhysRevLett.119.161101
  [arXiv:1710.05832 [gr-qc]].

\bibitem{Monitor:2017mdv} B.~P.~Abbott {\it et al.} [LIGO Scientific and Virgo and Fermi-GBM and INTEGRAL Collaborations],
  Astrophys.\ J.\  {\bf 848}, no. 2, L13 (2017)
  doi:10.3847/2041-8213/aa920c
  [arXiv:1710.05834 [astro-ph.HE]].

\bibitem{barrow1987lensing} J.D. Barrow and K. Subramanian,
Nature, {\bf 327}, 375 (1987).

\bibitem{Mena:2006ym} O.~Mena, I.~Mocioiu and C.~Quigg,
  Astropart.\ Phys.\  {\bf 28}, 348 (2007)
  [astro-ph/0610918].

\bibitem{Eiroa:2008ks} E.~F.~Eiroa and G.~E.~Romero,
  Phys.\ Lett.\ B {\bf 663}, 377 (2008)

\bibitem{Fan:2016swi} X.~L.~Fan, K.~Liao, M.~Biesiada, A.~Piorkowska-Kurpas and Z.~H.~Zhu,
  Phys.\ Rev.\ Lett.\  {\bf 118}, no. 9, 091102 (2017)

\bibitem{Wei:2017emo} J.~J.~Wei and X.~F.~Wu,
  Mon.\ Not.\ Roy.\ Astron.\ Soc.\  {\bf 472}, no. 3, 2906 (2017)

\bibitem{Yang:2018bdf} T.~Yang, B.~Hu, R.~G.~Cai and B.~Wang,
  Astrophys.\ J.\  {\bf 880}, 50 (2019)

\bibitem{Tanabashi:2018oca} M.~Tanabashi {\it et al.} [Particle Data Group],
  Phys.\ Rev.\ D {\bf 98}, no. 3, 030001 (2018).

\bibitem{Sakstein:2017xjx} J.~Sakstein and B.~Jain,
  Phys.\ Rev.\ Lett.\  {\bf 119}, no. 25, 251303 (2017)

\bibitem{Baker:2017hug} T.~Baker, E.~Bellini, P.~G.~Ferreira, M.~Lagos, J.~Noller and I.~Sawicki,
  Phys.\ Rev.\ Lett.\  {\bf 119}, no. 25, 251301 (2017)

\bibitem{Jia:2015zon} X.~Liu, J.~Jia and N.~Yang,
  Class.\ Quant.\ Grav.\  {\bf 33}, no. 17, 175014 (2016)

\bibitem{Pang:2018jpm} X.~Pang and J.~Jia,
  Class.\ Quant.\ Grav.\  {\bf 36}, no. 6, 065012 (2019)

\bibitem{Gibbons:2008rj} G.~W.~Gibbons and M.~C.~Werner,
  Class.\ Quant.\ Grav.\  {\bf 25}, 235009 (2008)

\bibitem{Ishihara:2016vdc} A.~Ishihara, Y.~Suzuki, T.~Ono, T.~Kitamura and H.~Asada,
  Phys.\ Rev.\ D {\bf 94}, no. 8, 084015 (2016)

\bibitem{Arakida:2017hrm} H.~Arakida,
  Gen.\ Rel.\ Grav.\  {\bf 50}, no. 5, 48 (2018)
  doi:10.1007/s10714-018-2368-2
  [arXiv:1708.04011 [gr-qc]].

\bibitem{Crisnejo:2018uyn} G.~Crisnejo and E.~Gallo,
  Phys.\ Rev.\ D {\bf 97}, no. 12, 124016 (2018)

\bibitem{Jusufi:2018jof} K.~Jusufi, A.~Övgün, J.~Saavedra, Y.~Vásquez and P.~A.~González,
  Phys.\ Rev.\ D {\bf 97}, no. 12, 124024 (2018)

\bibitem{Ovgun:2018tua} A.~Övgün, İ.~Sakallı and J.~Saavedra,
  JCAP {\bf 1810}, 041 (2018)
  doi:10.1088/1475-7516/2018/10/041
  [arXiv:1807.00388 [gr-qc]].

\bibitem{Kumaran:2019qqp} Y.~Kumaran and A.~Övgün,
  Chin.\ Phys.\ C {\bf 44}, 025101 (2020)
  doi:10.1088/1674-1137/44/2/025101
  [arXiv:1905.11710 [gr-qc]].

\bibitem{Zhu:2019ura} T.~Zhu, Q.~Wu, M.~Jamil and K.~Jusufi,
  Phys.\ Rev.\ D {\bf 100}, no. 4, 044055 (2019)
  doi:10.1103/PhysRevD.100.044055
  [arXiv:1906.05673 [gr-qc]].

\bibitem{Ovgun:2019qzc} A.~Övgün, İ.~Sakallı and J.~Saavedra,
  arXiv:1908.04261 [gr-qc].

\bibitem{Javed:2019kon} W.~Javed, J.~Abbas and A.~Övgün,
  Eur.\ Phys.\ J.\ C {\bf 79}, no. 8, 694 (2019)
  doi:10.1140/epjc/s10052-019-7208-3
  [arXiv:1908.09632 [physics.gen-ph]].

\bibitem{Li:2019qyb} Z.~Li and J.~Jia,
  Eur.\ Phys.\ J.\ C {\bf 80}, no. 2, 157 (2020)
  doi:10.1140/epjc/s10052-020-7665-8
  [arXiv:1912.05194 [gr-qc]].

\bibitem{Ono:2019hkw} T.~Ono and H.~Asada,
  Universe {\bf 5}, no. 11, 218 (2019)
  Some 2nd order terms in Eq. \eqref{eq:alphalbsas} were not given in this reference.

\bibitem{Sloane:1978ne} A.~Sloane,
  Austral.\ J.\ Phys.\  {\bf 31}, 427 (1978).

\bibitem{Ono:2017pie} T.~Ono, A.~Ishihara and H.~Asada,
  Phys.\ Rev.\ D {\bf 96}, no. 10, 104037 (2017)
  doi:10.1103/PhysRevD.96.104037
  [arXiv:1704.05615 [gr-qc]].

\bibitem{Jia:2020dap} J.~Jia,
[arxiv:2001.02038 [gr-qc]].

\bibitem{bk:hobson} Hobson,M.P., Efstathiou, G.P., Lasenby, A.N.: {\it General Relativity: An Introduction for Physicists}, Sec. 18.3, p. 503. Cambridge University Press, Cambridge (2006)

\bibitem{bk:inttable} I.S.Gradshteyn, and I.M.Ryzhik, \emph{Table of Integrals, Series, and Products}, 8th ed. Academic Press (2014), p. 152.

\bibitem{Bardeen:1973} J.~M.~Bardeen, Rapidly rotating stars, disks, and black holes, {\it Black Holes}, ed. C. DeWitt, B. S. DeWitt, Gordon and Breach Science Publishers, (1973)

\bibitem{Duan:2020tsq} Y.~Duan, W.~Hu, K.~Huang and J.~Jia,
  arXiv:2001.03777 [gr-qc].

\bibitem{Misner:1974qy} C.~W.~Misner, K.~S.~Thorne and J.~A.~Wheeler,
  San Francisco 1973, 1279p

\bibitem{suppmat} See Supplement Material [link] for deflection angle in equatorial plane of KN spacetime to 9-th order,

\bibitem{sgrapara} Andreas Eckart et al.,
PoS (APCS2018) 048

\bibitem{Weinberg:1972kfs} S.~Weinberg,
  {\it Gravitation and Cosmology: Principles and Applications of the General Theory of Relativity}, John Wiley \& Sons (1972). ($d_1$ however in general is always $2m$.)

\bibitem{Butcher:2016yrs} L.~M.~Butcher,
  Phys.\ Rev.\ D {\bf 94}, no. 8, 083011 (2016)

\end{thebibliography}
\end{document}